\documentclass{article} 
\usepackage{graphicx}
\usepackage{amssymb}
\usepackage{amsthm}
\usepackage{amsmath}
 
\newtheorem{Theorem}{Theorem}
\newtheorem{Remark}{Remark}

\title{On a coupled time-dependent SIR models fitting with New York and New-Jersey states COVID-19 data. }

\author{Benjamin Ambrosio, M.A. Aziz-Alaoui\\
Normandie Univ, UNIHAVRE, LMAH, FR-CNRS-3335,\\
 ISCN, 76600 Le Havre, France\\
 benjamin.ambrosio@univ-lehavre.fr\\
}

\begin{document}
\maketitle

\abstract{This article describes a simple Susceptible Infected Recovered (SIR) model fitting with COVID-19 data for the month of march 2020 in New York (NY) state. The model is a classical SIR, but is non-autonomous; the rate of susceptible people becoming  infected is adjusted over time in order to fit the available data. The death rate is also secondarily adjusted. Our fitting is made under the assumption that due to limiting number of tests, a large part of the infected population has not been tested positive. In the last part, we extend the model to take into account the daily fluxes between New Jersey (NJ) and NY states and fit the data for both states. Our simple model fits the available data, and illustrates typical dynamics of the disease: exponential increase, apex and decrease. The model highlights a decrease in the transmission rate over the period which gives a quantitative  illustration   about how lockdown policies reduce the spread of the pandemic. The coupled model with NY and NJ states shows a wave in NJ following the NY wave, illustrating the mechanism of spread   from one attractive hot spot to its neighbor. }
\section{Introduction and model}
From a scientific perspective, the COVID-19 pandemic has highlighted the crucial role of mathematical and statistical models in providing guidance for health policies. Expressions such as "flatten the curve", the "apex", the "plateau"  have been widely heard in medias and employed by decision makers to explain their choices regarding rules and policies during this critical period. In this short article, we first introduce a simple SIR model, in which we  adjust  a key parameter $k$ standing for a control on the Susceptible-Infected rate, and secondarily the death rate, in order to fit the data of the pandemic in NY state in March 2020, and provide predictions for a near future. Then, we add a node in the model to take into account the daily fluxes between NY and NJ states. Note that these two close states, are, up to the day of redaction of this article, the more severe  hit by the pandemic in United States. Of course, the coupling may be extended to other states. However, in this article, we restrict ourselves to NY and NJ. Accordingly, the main key points of this article are that, 1) it highlights the dynamics and epidemiological characteristics which have been discussed in press and health policies; It highlights qualitatively how lockdown policies have decreased the spread of the virus and provides prediction and explanation of an upcoming apex,  2) it fits real data provided for the New York state and 3) it fits the data of NJ state by considering  coupled equations taking into account the daily fluxes between NY and NJ. This provides a quantitative visualization of how the virus may spread from an attractive hot spot (New York City in NY state) towards close states trough the daily fluxes of commuters.

    We especially focus on fitting the total number of cases tested positive for COVID-19 as well as the number of deaths in both NY and NJ states. We also give insights in prediction  of the number of people needing hospitalization in NY state. 

SIR models are very classic in literature. For some reader's convenience, we mention here some contextual elements and references. 
The simplest classical SIR model is the Kermack–McKendrick (KMcK)  model which goes back to 1927, see \cite{Ker-1927,BookMurray}. It writes 
 \begin{equation}\label{eq:SIR-0}
(KMcK)\left\{
\begin{array}{rcl}
S_t &=&-kSI\\

I_t  &=&kIS-rI \\
R_t  &=& rI\\
\end{array}
\right.
\end{equation}
In this original (KMcK) model, the population splits into three classes. The class $S$ stands for susceptible, who can catch
the disease, $I$ stands for  infective, who have the disease and can transmit it and $R$ stands for  the removed, namely, those who have or have had the disease but not transmit it anymore. Note that our terminology is slightly different as explained below. In \eqref{eq:SIR-0}, the dynamics follow the scheme
\[S\rightarrow I\rightarrow R\]
with respective transfer rates between classes of $k$ and $r$.
  
  SIR type models and more generally, mathematical models of epidemics have in fact a significant history. We refer to \cite{BookMurray} for a textbook on these models and a brief history of epidemics. Models have  become more sophisticated and may include more compartments such as exposed, infective asymptomatic, infective with symptoms, and also reservoirs such as bats, and include stochastic dynamics. Recently, SIR type models have been widely used  in the context of the COVID-19 pandemic, to model the spread all around the world, see for example \cite{Hui2019,Lin2020,Chen2020,Kun2020,Liu2020,Liu2020-2,Tang2020} and reference therein cited.   Here are also some examples of references for SIR models in other epidemic diseases: Dengue \cite{Mis-2017},  Chikungunya \cite{Mou2011,Mou2012} and Ebola \cite{Thompson2019}. See also references therein cited.

In the present article, we first consider the following model 
 \begin{equation}\label{eq:SIR}
\left\{
\begin{array}{rcl}
S_t &=&-k(t)\frac{S}{I+S+R}I\\

I_t  &=&k(t)\frac{S}{I+S+R}I-r(t)I-d(t)I \\
R_t  &=& r(t)I\\
\end{array}
\right.
\end{equation}
This simple model has classically three classes: $S$ for susceptible, $I$ for infected and $R$ for recovered. Specifically, the class $I$ is intended to represent all the people who bear effectively the virus at a given time, and can transmit it if in contact with other people. It includes all infected people with or without symptoms, reported or not. 
There are some differences with equation \eqref{eq:SIR-0}. First, it includes a death rate $d(t)$. And even tough, the number of deaths does not appear explicitly as a variable, it is simply given by the integral $\int_0^td(u)I(u)du$.  Also note that in expression
\[S_t=-k(t)\frac{S}{S+I+R}I,\]the rate of contamination from $S$ to $I$ is proportional to the proportion of susceptible  ($S$) in the whole population ($S+I+R$). This is a classical expression standing for the fact that the probability for each individual in the $I$ class to spread the virus among the class $S$ is proportional to the portion of $S$ in the whole population, see for example \cite{Gao-1992,Yos-2007} and references therein cited. This rate is corrected by a crucial  coefficient $k(t)$ which is intended to fit the real transfer rate and which contains the effects inherent to the properties of the virus (for example change of propagation rate due to genetic mutation of the virus) or to specific policies (like quarantine, social distancing, lockdown...). This time dependence  allows to adjust the dynamics to fit the data. This  is a specificity of our model and turns it into a non-autonomous equation. This time dependence of $k$ is obviously  relevant in our model since the rate of transfer from $S$ to $R$ is the main target of health policies and is subsequently subject to vary over time. Secondarily, we also allow the death rate to vary. Many internal or external factors may affect the death rate among which are concomitant lethal disease, temperature, hospital conditions... More significantly, one has to note that the rate transfers considered here are  instantaneous transfer between compartments, and the function $d(t)$ is different from the Case Fatality Rate (CFR). Recall that the CFR is the death rate per confirmed case over a given period of time, and is a typical indicator for death rate. In South Korea, the country which led the highest number of tests, it has been reported to be of $1$ percent, see \cite{Lac-2020}. In China as of February 2020, this rate varied between 3.8 in the region of Wuhan and 0.7 in others regions, see \cite{who-1} and also \cite{Bau-2020}. Using tables 1 and 2 gives a CFR in NY from March 1 to April 1 of $\frac{1941}{83804}\simeq 2.3$ per cent. Since our model fits the data, by definition, it fits the CFR. At the end of the epidemic, if the whole number of people that contracted the virus would be tested positive the CFR would provide the probability to die for an individual  who catches the virus. However, during a growing phase such as the month of March considered here, the CFR has large variations. Moreover, there is a  delay between the time a person is tested positive and the time he dies. The
time between symptom onset and
death has been reported to range from about 2 to
8 weeks, see \cite{Bau-2020}. The typical average being 23 days according to \cite{Lac-2020}.  One could for example introduce a time-integral death expression $\int_{0}^t\tilde{d}(u,t)I(u)du$ to take into account these informations. However, the time-window considered here is short and corresponds to the beginning of reported cases in NY. Furthermore, in this short article, we wanted to focus on a simple model able to fit data, highlight relevant dynamics and provide estimations.  
Since, a person in the $I$ compartment, will either recover or die, above remarks on the function $d$ hold for the coefficient $r$. In the present work, for sake of simplicity, we have set the $r$ coefficient to the constant value $0.64$. This is a simplification which is classical in SIR models.  Note that setting  the coefficient $r$ to a constant value is equivalent to assume that people recovering between times $t-1$ and $t$, which is given by $R(t)-R(t-1)$ would also write $r\int_{t-1}^tI(u)du$ according to equations \eqref{eq:SIR}. A more meaningful expression for  $R(t)-R(t-1)$ would be $\int_{0}^t\tilde{r}(u,t)I(u)du$ standing to the fact that people recovering between time $t-1$ and $t$ have been infected from a period ranging from $0$ to $t$, with a transfer rate given by      $\tilde{r}(u,t)$. This would lead to the equation $ r=\frac{\int_{0}^t\tilde{r}(u,t)I(u)du}{\int_{t-1}^tI(u)du}$.
Once $r$ is set to a constant value, to fit data over a reasonable period, numerical tries provide a unique choice for constants  $k$ and $d$. It was still possible to use different values of $r$ to fit the data. However, different values of $r$ result in different dynamics over time, and notably different time for apex. Our choice of $r=0.64$ was made to provide dynamics that seem relevant to us regrading the timely dynamics beyond the data. In particular, too smaller of  values for $r$ would provide dynamics with a late apex and less relevant regarding the timely  effects of the disease. Other studies have considered models with a constant $r$, with different values. See for example \cite{Roq-2020} and references therein cited.   Remarkably, varying $r$, and make it depend on time, provides some freedom to later fit data over a longer period of time, taking in account the end of the first wave in NY.

Upon the above discussions, our strategy is  rather simple: set $r$ to a constant value. Then choose a constant $k$ which fits the data of positive cases during a period of time. Next, choose a constant $d$ to fit deaths data for the same period of time. Note that since the parameter $k$ has much more effect than the small parameter $d$, this procedure is possible and efficient. After, repeat the procedure over a subsequent period. The overall procedure results in a constant function $r$ and two piecewise constant functions $k(t)$ and $d(t)$. For our model and the given data, the procedure was efficient allowing to provide these choices by successive tries. Note that this could be done automatically by cooking an algorithm to set the parameters following these guidelines.   Note finally that our assumptions do not a include birth rate for the susceptible population but rather focus on the short time effects of the disease. 
\section{Numerical Simulations, Data and Dynamics}
\subsection{Fitting the total number of infected people and the number of deaths}
Data from total infected people and total number of deaths  in New York state is available at the New York times web site, see \cite{NYTimes2020}. We have downloaded the data  from there from March 1 to April 1, which makes 32 days.  We have reported the  number of total cases in table 1 and the number of deaths in table 2.\\
For numerical simulations of equation \eqref{eq:SIR}, the parameter $r$ was set to $0.64$. Then, in order to fit the data of the total number of infected people, we chose:
\begin{equation}
\label{eq:k}
k(t)=\left\{
\begin{array}{rcl}
k_1=1.057&\mbox{ if }& 0\leq t< 21 \\
k_2=0.9&\mbox{ if }& 21\leq t<24 \\
k_3=0.67&\mbox{ if }& 24\leq t< 27 \\
k_4=0.71&\mbox{ if }& 27\leq t\leq 32 
\end{array}
\right.
\end{equation}
and to fit the total number of deaths, we chose:
\begin{equation}
\label{eq:d}
d(t)=\left\{
\begin{array}{rcl}
d_1=0.0016&\mbox{ if }& 0\leq t<21 \\
d_2=0.00232&\mbox{ if }& 21\leq t<24 \\
d_3=0.00232&\mbox{ if }& 24\leq t<27 \\
d_4=0.0068&\mbox{ if }& 27\leq t\leq 32\\ 
\end{array}
\right.
\end{equation}
Initial conditions were set to 
\[(19453556,5,0),\]
19453556 being the number of people living in NY state in 2019 according to USA Census bureau. Regarding reported data in tables 1 and 2, as a matter of fact, not all infected people in March 2020  were tested. To take into account this fact in our model, we adjusted the parameters to fit the quantity $0.2\times(I+R)$ from the model,  with the total number of infected people (data in table 1) minus the total number of deaths (data in table 2), $i.e.$ we assume here that only 20 percent of living people that have contracted the virus has been tested positive. Note that shortly after the first submission of this work, NY state reported the result of a random antibody test over a sample of $3000$ individuals. Statewide, 13.9 percent of people were tested positive, ranging form 21.2 percent in New York City (NYC), to 3.6 percent in upstate counties. See \cite{Cuo-2020-2,Got-2020-2}. Comparing with data from \cite{NYTimes2020}, this gives a ratio of approximately 10; only about 10 percent of people having caught the virus in NY would have been tested positive.    Statistical estimates of related ratio, depending on the number of estimates provide a range of $[5,3000]$, see \cite{IHM-2020-1}. Simulation of system \eqref{eq:SIR}, and comparison with data resulting from tables 1 and 2  are plotted in Figure 1-a. In Figure 1-b, we plotted the total number of deaths from the model and compared it to the data. In Figure 1-c, we have plotted four curves corresponding to various $I(t)$ for different simulations of \eqref{eq:SIR}. 
\begin{itemize}
\item The curve $I_1(t)$ in red corresponds to the simulation of \eqref{eq:SIR} with $k(t)=k_1=1.057$ and $d(t)=d_1=0.0016$ for all time.
\item  The curve $I_2(t)$ in green corresponds to the simulation of \eqref{eq:SIR} with $k(t)=k_1=1.057$ and $d(t)=d_1=0.016$ for $0\leq t\leq 21$, and $k(t)=k_2=0.9$ and $d(t)=d_2=0.00232$ for $ 21 \leq t< 24$.
\item  The curve $I_3(t)$  in pink corresponds to the simulation of \eqref{eq:SIR} with $k(t)$ as given in \eqref{eq:k}, \textit{i.e.} $k(t)=k_i$ and $d(t)=d_i$, $t \in [t_{i-1},t_{i})$, $i\in \{1,...,4\}$ with  $t_0=0, t_1=21, t_2=24, t_3=27, t_4=32$.  
\end{itemize}
This panel illustrates how the health policies  flatten the curve.  In Figure 1-d, we have again plotted the solution $I(t)$ for $k(t)$ as in \eqref{eq:k}, for a longer period. 
{ \tiny
\begin{flushleft}
\hspace{-2cm}
\begin{table}
\label{ta:1}
\begin{tabular}{|c|c|c|c|c|c|c|c|c|c|c|c|c||c|c|c|c|c|}
\hline
\textbf{Day}&3/1&3/2&3/3&3/4&3/5&3/6&3/7&3/8&3/9&3/10&3/11\\
\hline
\textbf{Number of cases}&1&1&2&11&22&44&89&106&142&173&217\\
\hline
\textbf{Day}&3/12&3/13&3/14&3/15&3/16&3/17&3/18&3/19&3/20&3/21&3/22\\
\hline
\textbf{Number of cases}&326&421&610&732&950&1374&2382&4152&7102&10356&15168\\
\hline
\textbf{Day}&3/23&3/24&3/25&3/26&3/27&3/28&3/29&3/30&3/31&4/1&\\
\hline
\textbf{Number of cases}&20875&25665&33066&38987&44635&53363&59568&67174&75832&83804& \\
\hline
\end{tabular}
\caption{Total number of cases reported in NY state from march 1 to April 1. See \cite{NYTimes2020}.}
\end{table}
\end{flushleft}
}
\begin{table}
\label{ta:2}
\begin{tabular}{|c|c|c|c|c|c|c|c|c|c|c|c|c||c|c|c|c|c|}
\hline
\textbf{Day}&3/1&3/2&3/3&3/4&3/5&3/6&3/7&3/8&3/9&3/10&3/11\\
\hline
\textbf{Number of deaths}&0&0&0&0&0&0&0&0&0&0&0\\
\hline
\textbf{Day}&3/12&3/13&3/14&3/15&3/16&3/17&3/18&3/19&3/20&3/21&3/22\\
\hline
\textbf{Number of deaths}&0&0&2&6&10&17&27&30&57&80&122\\
\hline
\textbf{Day}&3/23&3/24&3/25&3/26&3/27&3/28&3/29&3/30&3/31&4/1&\\
\hline
\textbf{Number of deaths}&159&218&325&432&535&782&965&1224&1550&1941& \\
\hline
\end{tabular}
\caption{Total number of deaths reported in NY state from march 1 to April 1. See \cite{NYTimes2020}.}
\end{table}

\begin{table}
\label{ta:3}
\begin{tabular}{|c|c|c|}
\hline
&$ k(t)$&$d(t)$\\
\hline
$I_1(t)$ &$k(t)=1.057$ & $d(t)=0.0016$\\
\hline
$I_2(t)$ &$1.057$ if $t <21$; $0.9$ otherwise & $0.0016$ if $t<21$; $0.00232$ otherwise\\
\hline
$I_3(t)$& $k_i$ if $t_{i-1} \leq t < t_i, i\in \{1,...,4\}$ & $d_i$ if $t_{i-1} \leq t < t_i, i\in \{1,...,4\}$\\
\hline
\end{tabular}
\caption{Summary of the values used in Figure 1-c to obtain the curves $I_1(t)$, $I_2(t)$ and $I_3(t)$. Recall that $k_1=1.057, k_2=0.9, k_3=0.67, k_4=0.71$, $d_1=0.0016, d_2=0.00232, d_3=0.00232, d_4=0.0068$, $t_0=0$, $t_1=21$, $t_2=24$, $t_3=27$ and $t_4=32$.}
\end{table}

\begin{table}
\label{ta:hosp}
\begin{tabular}{|c|c|c|c|c|c|c|c|c|c|c|}
\hline
\textbf{Day}&3/16&3/17&3/18&3/19&3/20&3/21&3/22&3/23&3/24&3/25\\
\hline
\textbf{Total number of hospitalized}&326&496&617&1042&1496&2043&2629&3343&4079&5327\\
\hline
\textbf{Day}&3/26&3/27&3/28&3/29&3/30&3/31&4/1& & & \\
\hline
\textbf{Total number of hospitalized}&6481&7328&8503&9517&10929&12226&13383& & &\\
\hline
\end{tabular}
\caption{Total number of daily current hospitalizations reported in NY state from March 16 to April 1. See \cite{IHM-2020-1,Cuo-2020-1}.}
\end{table}

\begin{figure}
\begin{center}
\includegraphics[width=6cm,height=5cm]{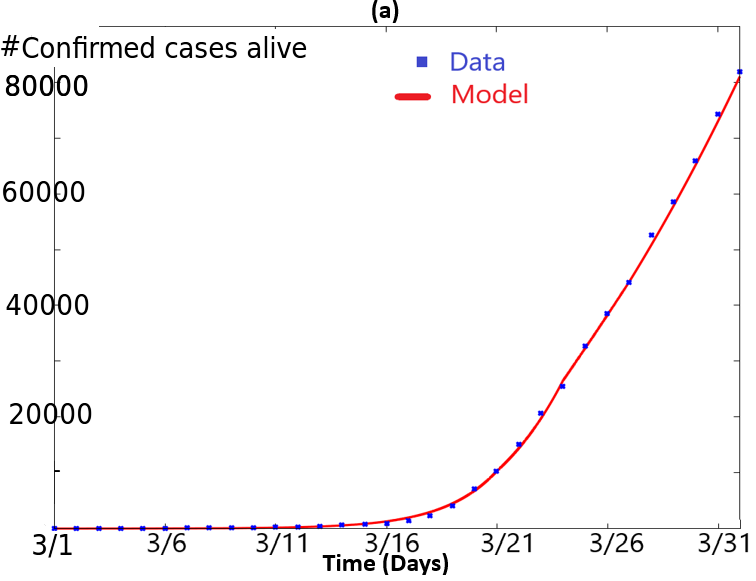}
\includegraphics[width=6cm,height=5cm]{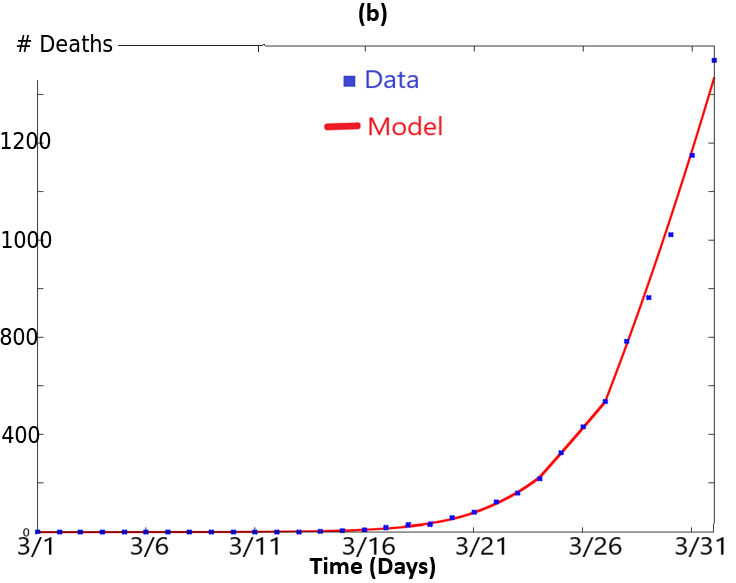}\\
\includegraphics[width=6cm,height=5cm]{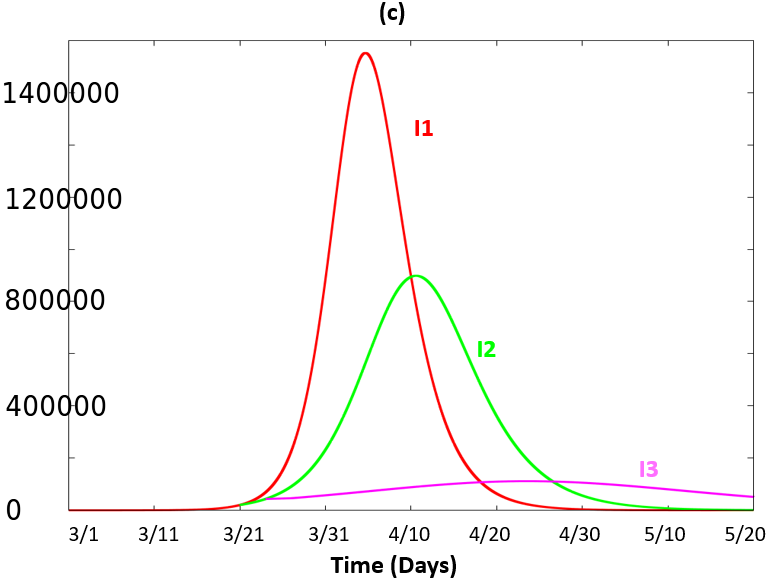}
\includegraphics[width=6cm,height=5cm]{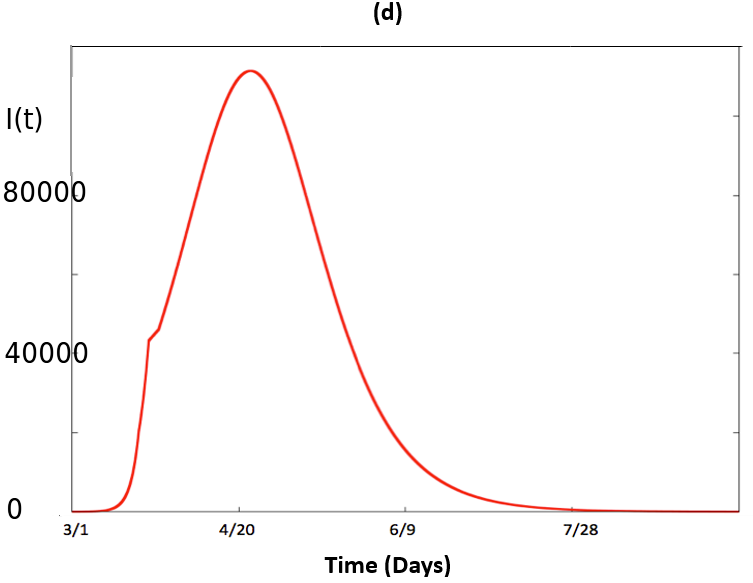}
\label{fig:1}
\end{center}
\caption{This figure illustrates the simulation of system \eqref{eq:SIR} and how it fits the data. In (a), we have plotted the quantity $0.2\times (I+R)$ as a function of time in red. The blue dots correspond to the data retrieved from \cite{NYTimes2020}. Analogously, in (b), we have plotted the quantity $\int_0^td(s)I(s)ds$, which represents the total number of deaths according with the model, as a function of time in red. The blue dots correspond to the data retrieved from \cite{NYTimes2020}. In (c), we have illustrated the quantity $I(t)$ corresponding to different values of $k(t),d(t)$:  the curve $I_1(t)$ in red corresponds to the simulation of \eqref{eq:SIR} with $k(t)=k_1$ and $d(t)=d_1$ for all time. The curve $I_2(t)$ in green corresponds to the simulation of \eqref{eq:SIR} with $k(t)=k_1$ and $d(t)=d_1$ for $0\leq t\leq 21$, and $k(t)=k_2$ and $d(t)=d_2$ for $t\geq 21$. The curve $I_3(t)$  in pink corresponds to the simulation of \eqref{eq:SIR} with $k(t)$ as given in \eqref{eq:k}, \textit{i.e.} $k(t)=k_i$ and $d(t)=d_i$, $t \in [t_{i-1},t_{i})$, $i\in \{1,...,4\}$ with  $t_0=0, t_1=21, t_2=24, t_3=27, t_4=32$.  It illustrates how the health policies  flatten the curve.  In (d), we have again plotted the solution $I(t)$ for $k(t)$ as in \eqref{eq:k}, for a longer period.  }
\end{figure}
\subsection{Fitting the total number of people at hospital}
Next, we have retrieved data corresponding to the number of people being at the hospital between March 16 and April 1. During this period, and after, NY state officially reported daily useful charts and statistics, on the local spread.  We have then computed an estimation of people being effectively at hospital at a given date.  We  denote $H(t_i),t_i\in\{16,17,...,32\}$ the total number of hospitalizations  at a given date. In order to fit the solution of \eqref{eq:SIR} with this data, we performed a linear regression between $(I(t_i)),t_i\in\{16,17,...,32\}$ and $(H(t_i)),t_i\in\{19,20,...,32\}$. The coefficients $a$ and $b$ of the linear regression were determined by  the least-square method, leading to the formulas:

\[a=\frac{<x,y>-n\bar{x}\bar{y}}{||x||^2-\bar{x}^2}\]
\[b=\bar{y}-a\bar{x}\]
with
\[x=(H(t_i),t_i\in\{16,17,...,32\}\]
\[y=(I(t_i)),t_i\in\{16,17,...,32\}\]
\[<x,y>=\sum_{i=1}^nx_iy_i,\,||x||^2=<x,x>,\, \bar{x}=\frac{1}{n}\sum_{i=1}^nx_i,\, \bar{y}=\frac{1}{n}\sum_{i=1}^ny_i, \, n=14.\]
 The result is plotted in Figure 2-a. It clearly shows a good approximation  by two distinct lines, corresponding to
\[a_1=10.3301, \,b_1=-1227.61\] 
and
\[a_2=1.91478, \,b_2=33560.3.\]

Then, a prevision of number of people needing hospitalization can be provided by the formula:
\[H(t)=\frac{I(t)-b_2}{a_2} \mbox{ if } I(t)\geq 41475.76,  H(t)=\frac{I(t)-b_1}{a_1} \mbox{ otherwise.} \]
The result is plotted in Figure 2-b.    
\begin{figure}
\begin{center}
\includegraphics[width=6cm,height=5cm]{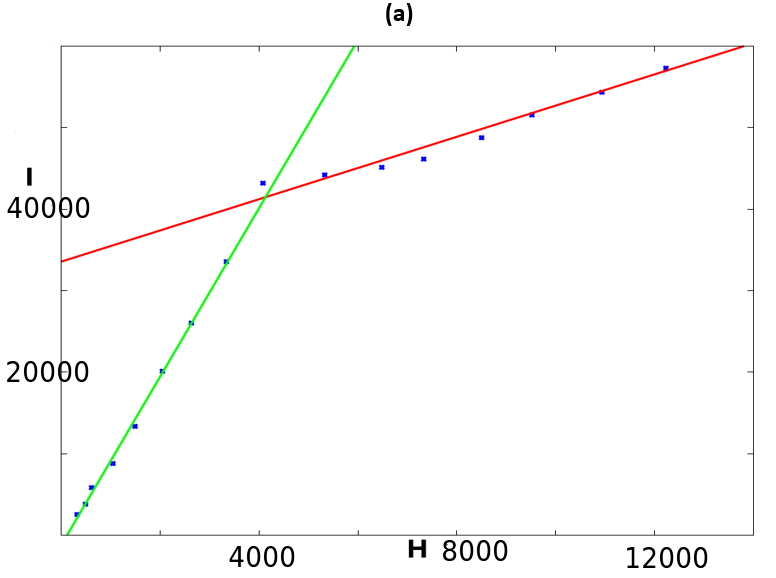}
\includegraphics[width=6cm,height=5cm]{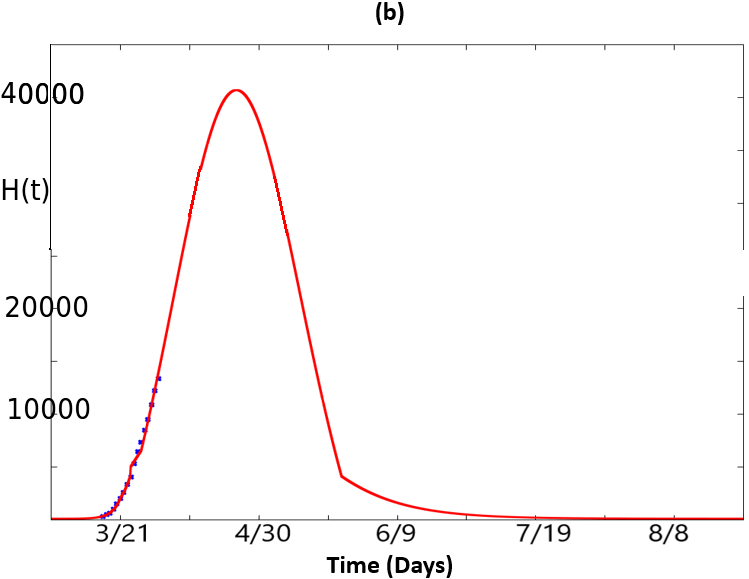}\\
\end{center}
\label{fig:2}
\caption{This figure illustrates how to provide an estimation for people needing hospitalization thanks to equation \eqref{eq:SIR} and statistical methods. Panel (a) illustrates an approximation of  $I(t_i),t_i\in\{16,17,...,32\}$ by a vector $(a_1H(t_i)+b_1),t_i\in\{16,17,...,23\}$ and another vector  $(a_2H(t_i)+b_2),t_i\in\{24,25,...,32\}$ where $a_j$ and $b_j$, $j\in\{1,2\}$ are the coefficients obtained thanks to the least-square method. Panel (b) then provides a prediction of people in need of hospitalization by plotting the quantity $H(t)=\frac{I(t)-b_2}{a_2}$ if $I(t)\geq 41475.76$,  $H(t)=\frac{I(t)-b_1}{a_1}$ otherwise. }
\end{figure}
\subsection{Dynamics}
Summing up the equations in \eqref{eq:SIR} and looking for stationary solutions yield the following theorem.
\begin{Theorem}
We assume that initial condition of system \eqref{eq:SIR} satisfies  $S(0)>0, I(0)>0$ and $R(0)=0$. Then for $t> 0$, all variables remain bounded and positive: there exists a positive constant $M<R(0)+I(0)$ such that for all $t>0$:
\[0<S(t)<M, 0<I(t)<M, 0<R(t)<M\] 
and 
\[S(t)+I(t)+R(t)\leq S(0)+I(0)-\int_0^td(s)I(s)ds.\]
Non-negative  stationary solutions of the system are given by $(\bar{S},0,\bar{R})$, with $\bar{S}\geq 0$ and $\bar{R}\geq 0$. Furthermore, $S(t)$ is decreasing, $R(t)$ is increasing and the variation of $I(t)$ is given by the sign of 
\begin{equation}
\label{eq:dynI}
k(t)\frac{S(t)}{S(t)+I(t)+R(t)}-r-d(t). 
\end{equation}  
\end{Theorem}
\begin{Remark}
 It is worth noting that this theorem, which proof is relatively straightforward provides two simple but relevant interpretations from the applicative point of view. The first thing is that the stationary solutions are given by $(\bar{S},0,\bar{R})$, with $\bar{S}\geq 0$ and $\bar{R}\geq 0$. This means that the stationary solutions, are all with $0$ infected but may take arbitrary non-negative values  (within  a bounded interval) for susceptible and recovered. This reflects the reality of the disappearance of the virus. The second thing we want to mention is that the variation of $I$ is given by the sign of expression \eqref{eq:dynI}. In particular, basically during a classical wave, this sign will be positive before the apex and negative after.  
\end{Remark}
\section{Two coupled SIR systems fitting COVID-19 for NY and NJ states}
It is estimated that around 400000 people used to  commute from NJ to NY  before the lockdown policies due to the pandemic. It is natural to integrate these effects
 in the model, since those commuters played the role of vector for virus before the lockdown. We therefore build a small network by coupling two nodes, representing NY (node 1) and NJ (node 2). To this end, we couple two copies of the local model \eqref{eq:SIR} 
to take into account the daily fluxes between those two states. This leads to  the following model:
\begin{equation}\label{eq:2-SIR}
\left\{
\begin{array}{rcl}
S_{1t} &=&-k_1(t)I_1\frac{S_1}{I_1+S_1+R_1}+l(t)\big{(}c^S_{12}(t)S_2-c^S_{21}(t)S_1\big{)}\\[1mm]
I_{1t}  &=&k_1(t)I_1\frac{S_1}{I_1+S_1+R_1}-rI_1-d_1(t)I_1 +l(t)\big{(}c^I_{12}(t)I_2-c^I_{21}(t)I_1\big{)}\\[1mm]
R_{1t}  &=& rI_1+l(t)\big{(}c^R_{12}(t)R_2-c^R_{21}(t)R_1\big{)}\\[2mm]
S_{2t} &=&-k_2(t)I_2\frac{S_2}{I_2+S_2+R_2}+l(t)\big{(}-c^S_{12}(t)S_2+c^S_{21}(t)S_1\big{)}\\[1mm]
I_{2t}  &=&k_2(t)I_2\frac{S_2}{I_2+S_2+R_2}-rI_2-d_2(t)I_2 +l(t)\big{(}-c^I_{12}(t)I_2+c^I_{21}(t)I_1\big{)}\\[1mm]
R_{2t}  &=& rI_2+l(t)\big{(}-c^R_{12}(t)R_2+c^R_{21}(t)R_1\big{)}\\
\end{array}
\right.
\end{equation}
The assumptions are analog to those given section 2, but we aim here to take into account  the daily fluxes between NJ and NY. For sake of simplicity, we consider that there is a flux of people coming from NJ to NY in the morning and returning home at night. The coupling functions, which are non-autonomous here are given by:
\[\begin{pmatrix}
l(t)\big{(}c^S_{12}(t)S_2-c^S_{21}(t)S_1\big{)}\\
l(t)\big{(}c^I_{12}(t)I_2-c^I_{21}(t)I_1\big{)}\\[1mm]
l(t)\big{(}c^R_{12}(t)R_2-c^R_{21}(t)R_1\big{)}\
\end{pmatrix} \] 
for node 1 and 
\[\begin{pmatrix}
l(t)\big{(}-c^S_{12}(t)S_2+c^S_{21}(t)S_1\big{)}\\
l(t)\big{(}-c^I_{12}(t)I_2+c^I_{21}(t)I_1\big{)}\\[1mm]
l(t)\big{(}-c^R_{12}(t)R_2+c^R_{21}(t)R_1\big{)}
\end{pmatrix} \] 
for node 2, where functions $+c_{ij}$ stand for the densities of population coming from mode $j$ and going into node $i$. In the remaining of the manuscript we assume that these functions $c_{ij}$ do not depend on $S,I$ and $R$ and drop the superscripts. Furthermore, we assume $c_{12}$ and $c_{21}$ to be periodic of period 1 (one day), with a  Gaussian profile and respective apex at 8:30 am and 6:30 pm. Not that the amplitude $c_{21}$ is multiplied by a  coefficient greater than 1. in comparison with $c_{12}$ to take in account the amount of population int NY and NJ. We want to  emphasize here the attractivity of NYC, and therefore assume that the fluxes are mainly from NJ to NY in the morning and form NY to NJ in the evening. The function $l(t)$  integrates policies of lockdown: after March 23, in the model, the daily fluxes are divided by $1000$. Therefore, the function $l(t)$ is a piecewise constant function given by:
\begin{equation}
 l(t):=\left\{
\begin{array}{rcl}
1& \mbox{ if } &0\leq t < 23\\[1mm]
 10^{-3} & \mbox{ if }&t \geq 23.
\end{array}
\right.
\end{equation}
Initial conditions were set to 
\[(19453556,5,0),\]
for node $1$ and 
\[(8882190,0,0),\]
for node 2. Note there are no infected cases at initial time in NJ.  This means that in our model, initial spread in NJ follows from infection in NY.
The same methods as in section 2 were used to fit the data for both NY and NJ for the network model \eqref{eq:2-SIR}. Illustrations are provided in Figure 3.
It shows  how the model fits the data. In Figure 3-a, we have plotted the quantity $0.2(I_2+R_2)$ as a function of time in red. Recall that, in the model $I_2+R_2$, represents the number of people in the population which has been infected by the virus and are still alive.  The blue dots correspond to the data retrieved from \cite{NYTimes2020} and plots the total number of infected minus the number of total  deaths. Analogously, in Figure 3-b we have plotted the quantity $\int_0^td_2(u)I_2(u)du$ as a function of time, in red. The blue dots correspond to the data retrieved from \cite{NYTimes2020}.  Finally, Figure 3-c illustrates $I_1t)$ and $I_2(t)$, which represent respectively the infected population in NY and NJ. It shows how the curve of NJ follows the curve in NY, with a small attenuation.  
\begin{figure}
\begin{center}
\includegraphics[width=6cm,height=5cm]{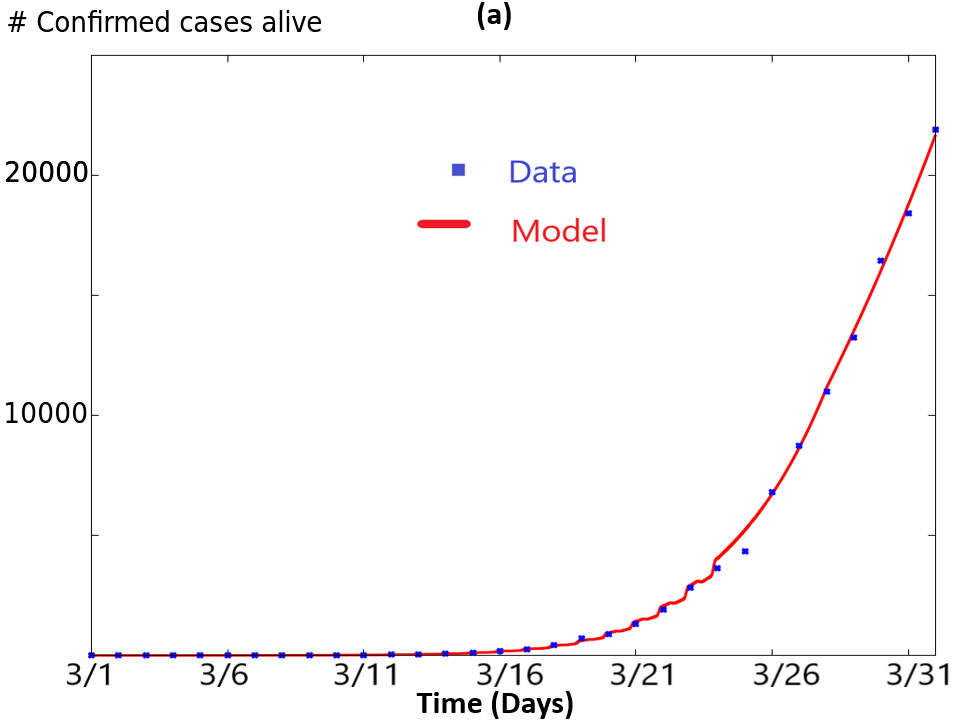}
\includegraphics[width=6cm,height=5cm]{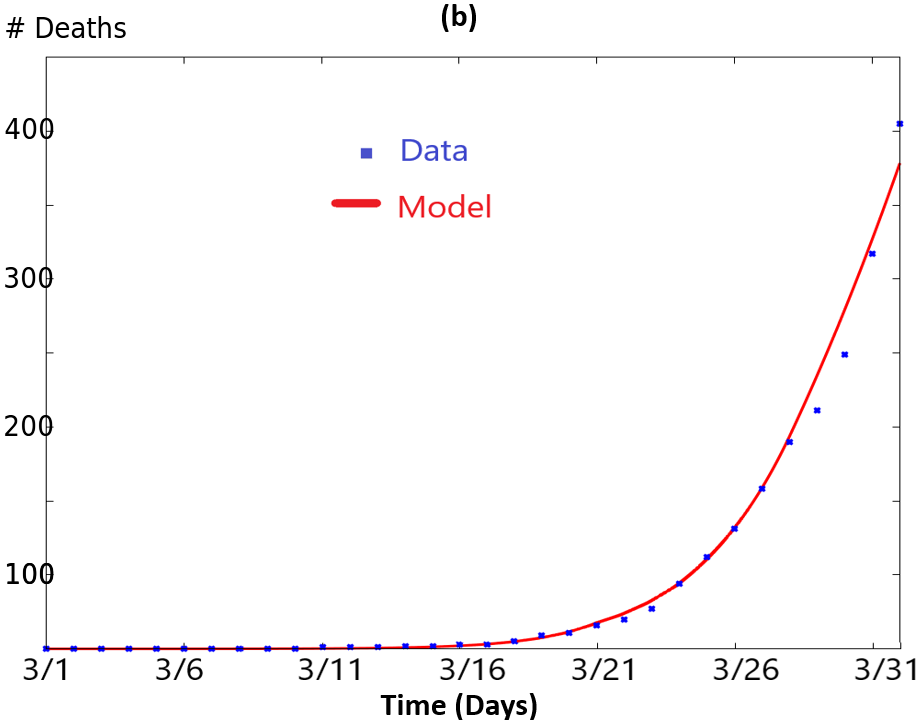}\\
\includegraphics[width=6.3cm,height=5.3cm]{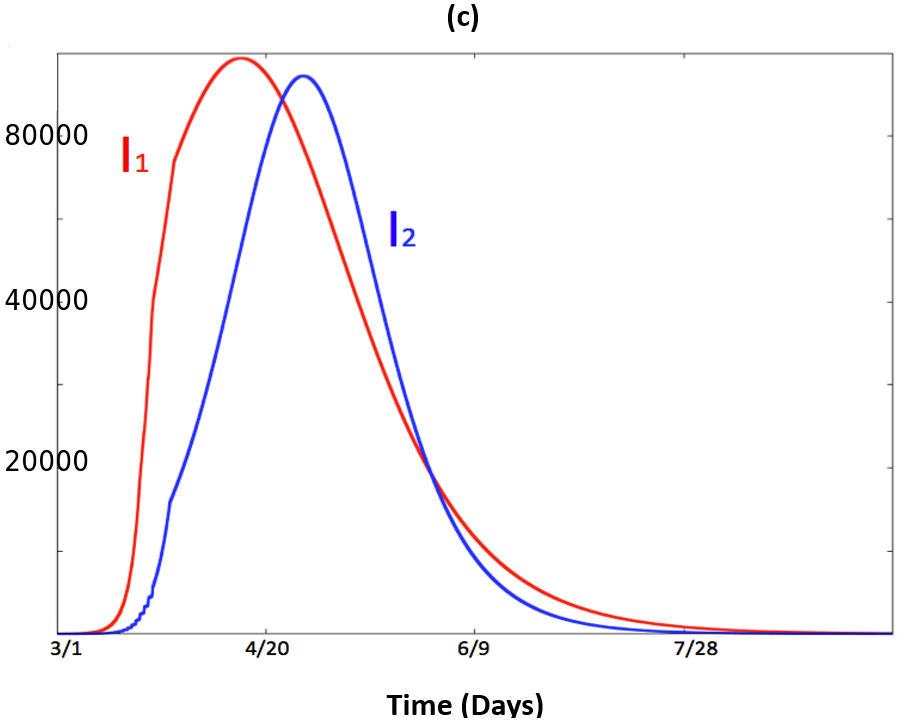}
\label{fig:3}
\end{center}
\caption{This figure illustrates the simulation of system \eqref{eq:2-SIR} and how it fits the data. In (a), we have plotted the quantity $0.2 (I_2+R_2)$ as a function of time in red. Recall that, in the model $I_2+R_2$, represents the number of people in the population which has been infected by the virus and are still alive. The blue dots correspond to the data retrieved from \cite{NYTimes2020} and plots the total number of infected minus the number of total  deaths. Analogously, in (b) we have plotted the quantity $\int_0^td2(u)I_2(u)du)$ as a function of time in red. The blue dots correspond to the total number of deaths in NJ according with data retrieved from \cite{NYTimes2020}. Panel (c) illustrates $I_1(t)$ and $I_2(t)$, which represent respectively the infected in NY and NJ.     }
\end{figure}
An analog theoretical result as in section 2 holds for solutions of \eqref{eq:2-SIR}.
\begin{Theorem}
We assume that initial condition of system \eqref{eq:2-SIR} satisfies  $S_1(0)>0, I_1(0)>0$, $R_2(0)=0$, $S_2(0)>0, I_2(0)=0$ and $R_1(0)=0$. Then for $t> 0$, all variables remain bounded and positive: there exists a positive constant $M<R(0)+I(0)$ such that for all $t>0$:
\[0<S_i(t)<M, 0<I_i(t)<M, 0<R_i(t)<M, i\in \{1,2\}\] 
and 
\[S_1(t)+I_1(t)+R_1(t)+S_2(t)+I_2(t)+R_2(t)\leq S_1(0)+I_1(0)+S_2(0)-\int_0^t(d_1(s)I_1(s)+d_2(s)I_2(s))ds.\]
For any non negative $S_1(0),R_1(0),S_2(0),R_2(0)$, the following  functions satisfying for $t>0$ 
\begin{equation*}
\begin{array}{c}
I_1(t)=I_2(t)=0\\
S_1(t)+S_2(t)=S_1(0)+S_2(0)\\
R_1(t)+R_2(t)=R_1(0)+R_2(0)\\
S_{1t}=-S_{2t}=l(t)\big{(}c_{12}(t)S_2-c_{21}(t)S_1\big{)}\\
R_{1t}=-R_{2t}=l(t)\big{(}c_{12}(t)R_2-c_{21}(t)R_1\big{)}
\end{array}
\end{equation*} 
are solutions of \eqref{eq:2-SIR}.  Furthermore, $S_1(t)+S_2(t)$ is decreasing, $R_1(t)+R_2(t)$ is increasing and $I_1(t)+I_2(t)$ is non increasing if and only if
\begin{equation}
\label{eq:DynI1I2}
I_1(t)k_1(t)\frac{S_1(t)}{S_1(t)+I_1(t)+R_1(t)}+I_2(t)k_2(t)\frac{S_2(t)}{S_2(t)+I_2(t)+R_2(t)}\leq I_1(t)(r+d_1(t))+I_2(t)(r+d_2(t)).  
\end{equation}
\end{Theorem}
\begin{Remark}
As in remark 1, we want to point out some interpretation of this theorem in the context of the pandemic. Here, we have described some solutions with free epidemic component ($I_(t)=I_2(t)=0$). Note that in this case however, $S_i$ and $R_i$ are not constant since they vary according to the fluxes between the two nodes. Again the last inequality, quantifies the idea that when the spread of the virus becomes lower than the death and recovery the infected population starts to decrease. The difference here is that we take into account the two nodes together. 
\end{Remark}
\section{Conclusion} 
   In this article, we have considered a simple non-autonomous SIR model to fit the data of COVID-19 in New York state. The model illustrates and quantifies how acting on the control $k(t)$ allows to flatten the curve of infected people over the time. From the model, using classical statistical methods, it is then possible to provide predictions of the number of people in needs of hospitalization. Lastly, we have fitted data from NJ state thanks to a coupled SIR model taking into account the daily fluxes between NJ and NY. It allows to predict similar dynamics in NY and NJ, with a delay and small attenuation. Note that, despite its simplicity, our model fits, as good as more sophisticated models, the available data during the growing phase of March 2020. In a forthcoming work, we aim to fit data with the model over a longer period . This would provide an accurate estimation of the different parameters.
\bibliographystyle{plain}
\bibliography{biblio}

\end{document}